\documentclass{mem}
\usepackage{natbib}
\usepackage{txfonts}
\usepackage{balance}
\usepackage[a4paper]{hyperref}
\idline{75}{282}
\begin{document}
\def\teff{$T\rm_{eff }$}
\def\kms{$\mathrm {km s}^{-1}$}

\title{
The Gaia Era: synergy between space missions and ground based surveys}

   \subtitle{}

\author{
A. \,Vallenari\inst{1},
R. \,Sordo,\inst{1}
          }

  \offprints{A. Vallenari}

\institute{
INAF--Osservatorio Astronomico, Vic. dell'Osservatorio 5
I-35122 Padova, Italy
\email{antonella.vallenari@oapd.inaf.it}
}

\authorrunning{Vallenari \& Sordo}

\titlerunning{The Gaia Era}

\abstract{The Gaia mission is expected to provide highly accurate astrometric, photometric, and spectroscopic measurements for about $10^9$ objects. Automated classification of detected sources is a key part of the data processing. Here a few aspects of the Gaia classification process are presented. 
Information from other surveys at longer wavelengths, and from follow-up ground based observations will be complementary to Gaia data especially at faint magnitudes, and will offer a great opportunity to understand our Galaxy.

}
\maketitle{}

\section{Introduction}

ESA's Gaia mission, to be launched in 2011, is meant to obtain accurate position, parallax and  proper motion for 10$^9$ object all over the sky, up to magnitude $G$=20 with an astrometric accuracy at the $\mu$arcsec level.
The low-dispersion spectroscopy obtained in the BP and RP passbands (330--1000 nm, resolution of 3--30 nm) will be used not only to correct the astrometry for color effects, but also to obtain a characterization of the sources themselves. The Radial Velocity Spectrograph (RVS, 840--890 nm, R$_{\rm P}$=11\,500) will measure radial velocities, with a precision of few km s$^{-1}$, up to $G$=16. Gaia will observe the whole sky for five years (plus a possible year extension) achieving a mean of $\sim$ 80 observations for each source. 
 The final catalog will be available in 2020, preceded by an early data release.  The data reduction is a great challenge: the size of Gaia related data will be about 10$^{15}$ bytes, while the final data delivered to the community would be of about 20 TB. The estimated computation size will be of the order of $10^{21}$ Flops \citep[see][]{mignard}.

\section{The classification of Gaia objects}

Gaia does not use a full sky input Catalog. 
However, in the early stages of the mission, the Guide Star Catalog \citep[][]{Lasker} (GSC-II)will be used as inputs for the  initial source list to support the identification of the  objects.
The GSC-II is constructed from the scanned images of the Palomar and UK Schmidt photographic survey digitized at the Space Telescope Science Institute. It makes use of Tycho-2 data as reference for the astrometric calibration. This is a good example of synergy between ground based surveys and spatial missions. Outside the Galactic plane the Catalog is complete down to R$_F \sim 20$, but stellar classification is reliable  at 90\% level at R$_F \sim 19.5$.
Coordinates are provided with a mean precision of $0.^{''}2$ --$ 0.^{''}28$ for about $9 \times 10^8$ objects.
The final Gaia catalog is expected to provide positions, parallaxes, proper motions, radial velocities, photometry in the two broad bands BP/RP, discrete classification of sources,  astrophysical parameters (APs) for single stars (i.e. T$_{\rm eff}$, $\log g$,...), and possibly the parametrization of special sources (galaxies, QSOs).
To avoid biases, and to built a reference frame for astrometry, Gaia will re-classify the observed objects. Such a large amount of data can be classified only in an automated way.
In the Gaia project, the classification algorithms   are based on both supervised and unsupervised methods  \citep[see][]{Coryn07}, first producing a discrete classification of the objects, i.e. dividing objects having higher probability of being stars, galaxies, and QSOs, then estimating its APs by comparison with a set of templates. Finally, the treatment of the outliers will relay on unsupervised methods.
The scientific community involved in Gaia is working to calculate extensive libraries of synthetic and observational templates with improved physics for all the classes of objects  to be used as training data for the classification algorithms.
In the classification task, Gaia data should be complemented by  external (i.e. non-Gaia) information. The simplest way to do this is via astrometric cross matching to existing catalogs. The most obvious candidates are the spectral and/or morphological classifications from SDSS and 2MASS, later also UKIDSS and PS1. FIRST could also be useful for the QSOs. This information could easily be introduced in a multi-component discrete classifier by means of the introduction of priors as pre-data estimate that a object belongs to a given class \citep[see][for a detailed description of the method]{Coryn08}. 

\section{Training data for object classification}

The Gaia object classification includes as well the determination of the APs of stars and possibly galaxies. As we state in the previous Section, supervised methods require the comparison with a set of templates, either observed or synthetic, as training data sets. While observational programs have already started to built a homogeneous  sample of stellar templates, it is clear that training data cannot be purely observational, since  a large variety and uniform coverage is requested  for the parameter distributions.
It turns out that high resolution and high quality synthetic libraries are of fundamental importance.
 New extensive calculations of sets of spectral stellar libraries with improved physics are on the way. They  cover the two Gaia spectral ranges: 300--1100 nm at 0.1 nm resolution, and 840--890 nm at 0.001 nm resolution. These new libraries  span a large range in atmospheric parameters, from super-metal-rich to very metal-poor stars, from cool stars to hot, from dwarfs to giant stars, with small steps in all parameters, typically $\Delta$\teff=250~K (for cool stars), $\Delta \log g$=0.5 dex, $\Delta$[Fe/H]=0.5 dex.  Depending on \teff, these libraries rely on MARCS \citep[F,G,K stars:][]{Gust08}, PHOENIX \citep[cool and C stars:][]{Brott05}, KURUCZ, TLUSTY  models including magnetic field, peculiar abundances,  mass loss \citep[A,B,Be,O stars:][]{Sordo06, Bouret08, Alva98, Koch05}.  WDA and WDB objects are included \citep{WD}. Those models are based on different assumptions: KURUCZ are LTE, plane-parallel, MARCS implement also spherical symmetry while PHOENIX and TLUSTY (hot stars) can calculate NLTE models both in plane-parallel mode and spherical symmetry \citep[see for a more detailed discussion][]{Gust08, Sordo2008}.
A comparable effort is carried on in the galaxy domain.
We remind that Gaia will extend the existing   surveys of galaxies (see or instance the SDSS  covering only a fifth of the sky) since it will be able to detect about $10^7$ unresolved galaxies down to G=20 covering the whole sky for the first time since photographic surveys (UK, ESO, Palomar Schmidt) of 30 years ago in a larger spectral range.
Large synthetic libraries of galaxy spectra covering the main Hubble types in the Gaia spectral range at a sampling of 1 nm or less are under construction \citep[see][]{Vivi}. At present, a library of about 3800 galaxy at zero redshift, and a second one of about 140,000 spectra at changing redshift are available.
Finally, QSOs synthetic and semi-empirical libraries \citep{Claeskens} are calculated.

\section{QSO classification: Gaia reference frame}
 
A high precision reference frame in Gaia is obviously mandatory to reach a high accuracy of 10~$\mu as$ in the astrometry.
With such a request, the astrometry must be self-calibrating and for this reason Gaia must observe a large number of quasars to define a  high precision  reference frame. This quasar sample must be very clean, showing only a low contamination by other objects. A probabilistic classifier is built to select objects with higher probability.  Care is paid to construct a pure sample of objects, discarding for instance  QSOs with low equivalent width emission lines which can be easily confused with F-G-K stars (4000-8000~K) having high extinction ($A_V \sim 8-10$).
Preliminary results  show that a pure sample of QSOs at 65\% level complete down to G=20 can be selected. This is adeguate to establish a reference frame for Gaia: Gaia will observe 500\,000 QSOs brighter than G=20, but only the objects with the most accurate positions ($G< 18$) will be used to built a reference frame. Following our preliminary estimates, a sample of 250\,000 can be recovered with no more than 13 contaminats \citep[see][]{Coryn08}. 
It is clear that the Gaia reference frame needs to be aligned with the International Celestial Reference Frame (ICRF) with the highest accuracy. The ICRF is based on VLBI positions of about 700 extragalactic radio sources. For this reason an observational program is initiated at the VLBI to identify suitable radio sources to align the two reference systems. At the moment, only a few objects can be useful to this purpose, either because they are not bright enough at optical wavelengths, or because they have extended emissions in the radio which precludes to reach the requested astrometric accuracy \citep{bourda}. 

\section{Gaia and complementary surveys}

Gaia will be of fundamental importance to study the Galactic structure at low latitudes: the position and the velocity of the OB stars con be measured without assuming rotation curve or extinction. The distances of OB stars at 4 Kpc with Av= 4 mag extinction will be derived with an accuracy of 13\%, space velocities with an accuracy of  a few Km/s. Fainter stars can be measured as well, giving important information about the mass distribution. However it should be noticed that at faint magnitudes (G $\sim$ 20), the expected accuracy is degraded and stellar APs cannot be reliably determined.
Once that a three-dimensional map of the Galactic plane is derived, and that distances and kinematics are known for all the star forming regions, we will be able to trace the disk and spiral structure of the Galaxy. The challenge will be to built dynamical theories to reproduce the density fields and the velocity fields. 
 To distinguish between m=2 and m=4 spiral arm structure recent simulations   find out that the potential needs to be known to 10\%,  the line-of-sight velocity accuracy needs to be better than 20 Km/s, distances shoud be known with uncertainties better than 30\% \citep[][]{Minchev2008}.
This is well  within the possibility limits of the Gaia survey.
However, two main problems should be reminded: the first is related  to the dust obscuration, which might hamper the determination of the star mass density, while the second is due to the confusion of the stars in the field of view (FOV). 
Concerning extinction, its knowledge  will become a limiting factor in the determination of the stellar luminosities and APs. The estimate made on the basis of G, BP, RP photometry only present some degree of degeneration: it is difficult to derive both the extinction and the extinction law for late type giants. Using RVS information, both parameters can be derived. In addition, the use diffuse interstellar bands (DIBs) are extinction tracers will be  explored. The strongest expected DIB in the surveyed range is at 862 nm. Its equivalent width well correlates  with the  interstellar reddening \citep[][]{Munari2008}. A large effort in going on in the Gaia community to ensure a proper determination of the extinction, testing and comparing different methods, including the use of infrared passbands in combination with Gaia passbands (Knude \& Lindstroem 2007). 
In addition, Gaia is confusion limited (BP/RP images of different stars are superposed on the FOV) when the total star density per transit (sum of the star number of both FOVs) is higher than 750,000 stars/sq deg.  This means that the central inner degrees  will not be well measured. Bright stars can probably still be recovered, but the precision will decrease. Simulations are still ongoing \citep{Marrese}. 
Infrared surveys are complementary to Gaia to understand the structure of the inner disk and deal with dust obscuration. Current astrometric data in the infrared are of poor quality, even if great improvements were made in the recent past (see 2MASS and DENIS). The UKIDSS will cover only a part of the sky (7000 sq deg), but it will be much deeper (K$\sim18-19$ on the Galactic plane, and K$\sim21$ at higher latitudes over 35 sq deg). Interferometric and adaptive optics astrometry can be very promising, but they can only  provide high precision relative astrometry in small fields. However, absolute parallaxes in small fields can be derived with high precision from these observations when enough suitable extragalactic reference sources are detected in the field of view.
Pan~STARRS and LSST  optical-near infrared surveys covering Northern and Southern sky respectively, are expected to observe $10^{10}$ stars down to magnitudes brighter than $24$ mag.  reaching an accuracy of about 3-25, 1-10 mas, respectively on the parallax determination. 
One of the space infrared survey which is foreseen in the near future is JASMINE, which is expected to cover the Bulge and the inner disk regions. JASMINE however, will not go very deep (z$<14$) observing about 10$^7$ stars with a precision of 0.01  mas on the parallaxes. All those surveys will be of great importance to map highly obscured regions where Gaia is not efficient,  although they cannot reach the same accuracy on astrometry.
Finally, since Gaia radial velocities will be measured only for $G<16$, spectroscopic ground based follow-up with 8m class telescopes need to be planned.

In conclusion, the Gaia mission will provide highly accurate astrometric, photometric, and spectroscopic measurements for a large sample of objects. The high quality of Gaia data, especially on  astrometry will not be reached by any of the planned surveys in the near future. Gaia data complemented by information coming from surveys at longer wavelengths, and from follow-up ground based observations will offer a great opportunity to unveil the formation and the structure  Galaxy.

\begin{acknowledgements}

This short review relies on the work of many members of the Gaia community, who are gratefully acknowledged for their valuable contribution to the project. 
\end{acknowledgements}

\bibliographystyle{aa}

\end{document}